\documentclass[structabstract]{aa}

\usepackage{graphicx}
\usepackage{natbib}
\usepackage{aas_macros}

\newcommand{\figwidth}{\columnwidth}
\newcommand{\Planck}{{\textit{Planck}}}

\title{Measuring Planck beams with planets}

\author{Kevin~M.~Huffenberger \inst{1,2,3}  \and Brendan~P.~Crill \inst{2,3} \and Andrew~E.~Lange \inst{2,3}\and Krzysztof~M.~G\'orski \inst{2,3}\and Charles~R.~Lawrence\inst{2,3}}

\institute{ 
University of Miami, Knight Physics Building, 1320 Campo Sano Dr., Coral Gables, FL 33124
\and
Jet Propulsion Laboratory, California Institute of Technology, 4800 Oak Grove Dr., Pasadena, CA 91109 
\and 
California Institute of Technology, 1200 E. California Blvd., Pasadena, CA 91125}

\abstract
{}
{Accurate measurement of the cosmic microwave background (CMB) anisotropy  
requires precise knowledge of the instrument beam.
We explore how well the \Planck\ beams will be determined from  
observations of planets, developing techniques that are also  
appropriate {for} other experiments.}
{We simulate planet observations with a \Planck-like scanning strategy, telescope beams, noise, and detector properties.  Then we employ both parametric and non-parametric techniques, reconstructing beams directly from the time-ordered data.  With a faithful parameterization of the beam shape, we can constrain certain detector properties, such as the time constants of the detectors, to high precision.  Alternatively, we decompose the beam using an orthogonal basis.  For both techniques, we characterize the errors in the beam reconstruction with Monte Carlo realizations.  For a simplified scanning strategy, we study the impact on estimation of the CMB power spectrum.  Finally, we explore the consequences for measuring cosmological parameters, focusing on the spectral index of primordial scalar perturbations, $n_s$.}
{The quality of the power spectrum measurement will be significantly influenced by the optical modeling of the telescope.  In {our most conservative} case, using no information about the optics except the measurement of planets, we find that a single transit of Jupiter across the focal plane will measure the beam {window} functions to better than 0.3\% for the channels at 100-217 GHz that are the most sensitive to the CMB.  Constraining the beam with optical modeling can lead to much higher quality reconstruction.}
{Depending on the optical modeling, the beam errors may be a significant contribution to the measurement systematics for $n_s$.}


\begin{document}

\maketitle

\section{Introduction}

Robust measurements of the cosmic microwave background (CMB) anisotropy, the source of  much of our understanding of the universe's contents, geometry, and primordial fluctuations, require detailed control over the systematics of the instrument.  The spatial response to a signal on the sky, known as the point-spread-function (PSF) or simply the telescope beam, is an important systematic effect because it smooths the anisotropy on the sky, damping high spatial frequencies in the angular power spectrum and washing out the encoded cosmological information. To recover the power spectrum, we face the challenging task of accurate beam reconstruction.

The release of the five-year results from the \textit{Wilkinson Microwave Anisotropy Probe} (WMAP) \citep{2009ApJS..180..225H}  highlights the issue's importance.  \citet{2009ApJS..180..246H} substantially refines the model of the instrument beam over the previous version, which is then folded into the power spectrum estimate \citep{2009ApJS..180..296N}.  The result is an increase in the five-year power spectrum over the three-year of 2 percent at the first acoustic peak and slightly more at smaller scales.  These changes are easily {visible} by eye when plotting the three year and five year spectra together, and outside the nominal error bars taken from the diagonal of the covariance matrix.  Because of the conservative treatment of beam errors in the three-year release likelihood method, the cosmological estimates fortunately do not change much---mostly manifesting as a $0.7 \sigma$ shift in the present-day amplitude of perturbations, $\sigma_8$ \citep{2007ApJS..170..377S, 2009ApJS..180..306D, 2009ApJS..180..330K}.

In this work, we examine the recently launched \Planck\ mission\footnote{\Planck\ (\emph{http://www.esa.int/Planck}) is a project of the European
Space Agency---ESA---with instruments provided by two scientific 
Consortia funded by ESA member states (in particular the lead countries: 
France and Italy) with contributions from NASA (USA), and
telescope reflectors provided in a collaboration between ESA and a 
scientific Consortium led and funded by Denmark.} \citep{Tauber_mission,2006astro.ph..4069T}, the next generation satellite to measure the CMB anisotropy.  To wring the full cosmological information from the observations, the proper calibration of the beam over a wide range of angular scales will prove crucial.  Because the sensitivity of the detectors alone would allow a cosmic variance limited measurement of the temperature power spectrum to high multipoles, determination of the physics at high spatial frequency will depend on the removal of systematics, in particular the quality of the beam reconstruction, especially since errors in the beam imprint errors on the power spectrum which are strongly correlated between multipoles.  

The beam error thus will affect a diverse range of science goals for the \Planck\ CMB maps and power spectra.
These include constraints on the early universe, in the measurement of the primordial spectrum's slope and running, the CMB damping tail, or any exotic physics at recombination \citep[e.g][]{2009MNRAS.398.1621C}.  In the later universe the high-$l$ spectrum affects constraints on the matter distribution from CMB lensing \citep[e.g.][]{2004NewA....9..687A,2003PhRvD..67l3507K,2003PhRvD..68h3002H,2005PhRvD..71h3008L,2006PhR...429....1L}, lensing-derived limits on neutrino masses  from CMB alone \citep{2003PhRvL..91x1301K,2005PhRvD..71d3001I} and in combination with other large scale structure data \citep{2006PhRvD..74l3005K,2008PhRvD..78h3535D}, information on cluster physics from the Sunyaev-Zeldovich (SZ) power spectrum \citep{2002MNRAS.336.1256K,2004MNRAS.352..993D,2007MNRAS.382.1697H}, and models of correlations in point source populations \citep[e.g.][]{2008A&A...478..685R}.  Finally, uncertainty in the beam shape adds error to cluster SZ and point-source flux measurements.
In addition, understanding \Planck's beam error is important for other experiments when forecasting cosmological performance based on \Planck\ prior parameter constraints.

Historically, CMB experiments have used a combination of optics calculations and planet measurements to work out the shape of the beam \citep[e.g.][among many others]{2003ApJS..148...39P,2003ApJS..148..527C,2006A&A...458..687M}.  Planets prove so useful because as bright, compact sources, they resemble $\delta$-function signal impulses.  In Sec.~\ref{sec:method}, we discuss \Planck's planet observations during the course of routine operations, our pipeline for simulating planet observations, and two methods for measuring the structure of the instrument beam: one in which we use significant prior information about the beam's shape, and another where we use very little.  We then interpret these beam reconstructions in terms of their effect on the CMB power spectrum.  In Sec.~\ref{sec:results}, we present the results of our computations, including detailed forecasts for the characterization of the beam reconstruction and uncertainties.  In Sec.~\ref{sec:cosmology}, we examine the impact on the scalar perturbation spectral index, $n_s$.  Finally, we conclude in Sec.~\ref{sec:conclusions}.

\section{Methods} \label{sec:method}

\subsection{Scan strategy and planet properties} \label{sec:planck_sees_planets}

The \textit{Low-Frequency Instrument} (LFI, \citealt{LFI_launchpaper}) and the \textit{High-Frequency Instrument} (HFI, \citealt{HFI_launchpaper}) of the \Planck\ spacecraft will observe the CMB from an orbit around the Earth-Sun Lagrange point, L2.  The spacecraft spins at $\sim 1$ rpm with the spin axis pointed roughly in the anti-Sun direction, sweeping beams (oriented $85^{\circ}$ from the spin axis) across the sky.  The spin axis is stepped along the ecliptic, about hourly, to keep the spin axis pointed away from the Sun.  The detailed strategy \citep{2005A&A...430..363D, Tauber_mission} modulates the spin-axis direction in a cycloid pattern, yielding virtually complete sky coverage in 7.5 months.

Since the anti-Sun direction and planet ephemerides may be predicted well in advance\footnote{Available online via the JPL HORIZONS system: \texttt{http://ssd.jpl.nasa.gov/?horizons}}, it is straightforward to estimate the times for planet observations by \Planck.
Because the exact observation time of a planet by an individual detector depends on the detector's position within the focal plane, the details of \Planck's orbit around L2, and constraints on the spin axis modulation, our estimates are correct to $\sim$ 1 week.
Planet brightness is determined by the orbital configuration during \Planck\ observation.    
Since the outer planets orbit the Sun more slowly than \Planck, they will be observed roughly once per sky survey.  {The orbital geometry required for observation (the planet is seen from L2 at $\pm 85^\circ$ from the anti-Sun ray) means the distance to the planet, and hence the brightness, is similar at every observation.}  The planets will be at high galactic latitude during their observation by \Planck\ in 2009 and 2010, so we have not included galactic emission here.

The data from a single observation of a planet will consist of a consecutive set of measurements for the duration of the focal plane's transit  over the planet.  Since planets lie near the ecliptic, the {detector pointings} in a single observation fall in stripes which are approximately perpendicular to the ecliptic.

\begin{figure}
  \begin{center}
    \includegraphics[width=0.49\figwidth]{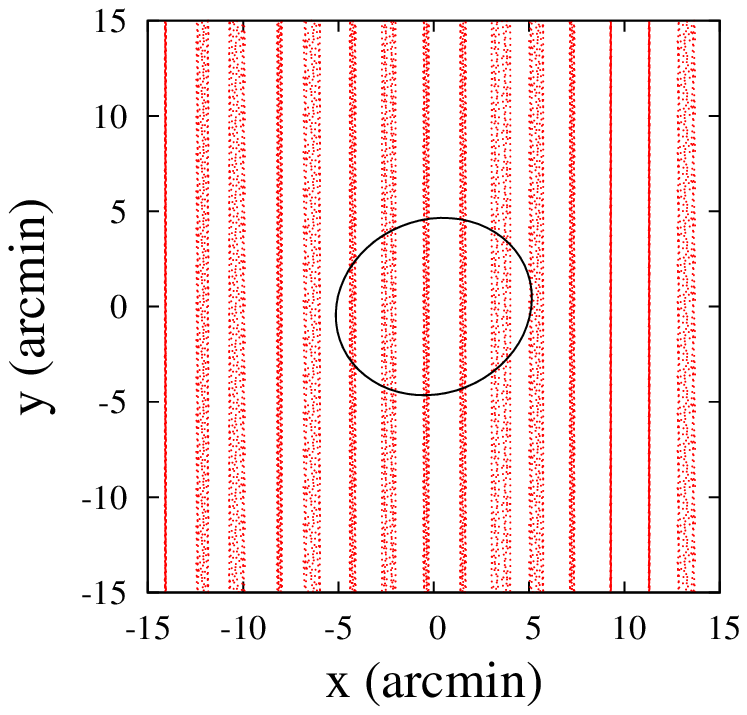}
    \includegraphics[width=0.49\figwidth]{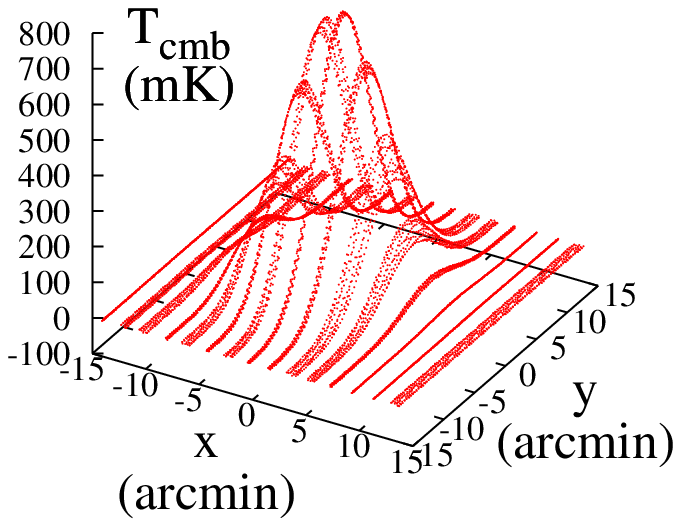}
  \end{center}
  \caption{Simulated Jupiter observation with a \Planck\ 100 GHz horn. Each point represent a single sample of the time-ordered data, and the coordinates are in the planet frame, with the x-axis in the cross-scan direction.  In the left panel, the telescope pointing at each sample is clearer, and the half-maximum curve is marked.}  \label{fig:pointing}
\end{figure}

Here we consider two realistic effects on the pointing of the spacecraft which reflect dynamics simulations from ESA \citep{Tauber_mission}.
First, every re-pointing will differ from the desired pointing by a small random error.  Second, the spacecraft spin axis nutates with a $\sim 6$ minute period  and an amplitude that may change after each re-pointing of the spacecraft.  The nutation of the spacecraft spin axis spreads samples along in cross-scan direction.  We based these effects on pointing simulations, but final values will be determined in flight.  Figure \ref{fig:pointing} shows a simulation of a Jupiter observation with \Planck-like pointing, using the pipeline developed for this work.

Estimates of the brightness temperatures for each planet in each     band are shown in Table \ref{tab:planet_brightness}.
Using angular diameters from the ephemerides, brightness, and assuming nominal beam sizes, we can calculate the beam dilution factors to find the signal seen by each  channel.    Jupiter and Saturn will be extremely bright in all the bands.  Mars, Uranus and Neptune will be seen with high signal to noise and will be useful for the main lobe of the beams (as may thousands of galactic and extra-galactic compact sources). 

\begin{table*}
\begin{center}
\begin{tabular}{lccccc} \hline \hline
Band  &\multicolumn{5}{c}{Peak temperature (mK)} \\
(GHz)	&	Jupiter$^1$	&	Saturn$^1$	&	Mars$^2$	&	Uranus$^3$	&	Neptune$^3$	\\
											
\hline										\\	
30	&	44.8	&	7.72	&	2.41	&	0.290	&	0.194	\\
44	&	90.6	&	15.3	&	4.75	&	0.564	&	0.319	\\
70	&	303	&	49.3	&	15.0	&	1.78	&	0.830	\\
100	&	754	&	121	&	38.1	&	4.29	&	1.73	\\
143	&	$1.86\times10^{3}$	&	299	&	91.8	&	9.12	&	3.66	\\
217	&	$7.58\times10^{3}$	&	$1.22\times10^{3}$	&	350	&	31.9	&	12.7	\\
353	&	$3.59\times10^{4}$	&	$5.60\times10^{3}$	&	$1.67\times10^{3}$	&	131	&	51.9	\\
545	&	$3.72\times10^{5}$	&	$6.50\times10^{4}$	&	$2.31\times10^{4}$	&	$1.59\times10^{3}$	&	630	\\
857	&	$3.31\times10^{7}$	&	$5.14\times10^{6}$	&	$1.89\times10^{6}$	&	$1.11\times10^{5}$	&	$4.43\times10^{4}$	\\
\\
\hline\hline

\end{tabular}

\end{center}
\caption{Model peak temperature (CMB thermodynamic units) in the \Planck\ beam for planets.  
References: (1) \citet{2007AstBu..62..285N}; (2)  \citet{1997ApJ...488L.161G,1976ApJ...210..250W}; (3) \citet{1993Icar..105..537G}.}
\label{tab:planet_brightness}
\end{table*}

\subsection{Beam fitting methods}

We have experimented with several methods to fit the beams.  Our original efforts concentrated on fitting beams to simulated, noisy maps of planet observations, a procedure which we ultimately found unsatisfactory.  We made maps two ways: binning on a rectangular grid after an offset subtraction (to remove long-time drifts) and onto a HEALPix grid\footnote{\texttt{http://healpix.jpl.nasa.gov}} \citep{2005ApJ...622..759G} using the destriping mapmaker \textit{Springtide} \citep[e.g. ][]{2009A&A...493..753A}, and attempted to fit beam parameters by minimizing $\chi^2$ over the pixels.  Initial fits gave rough beam parameters, but they were incorrect in detail due to the pixelization effects.  For the smaller beams, map pixels near bright planets contain large signal gradients (at the resolution of the \Planck\ CMB maps).  In principle increasing the map resolution solves this problem, but the number of pixels required approaches the number of time samples.  Therefore we abandoned map-domain fitting for time-domain fitting, the focus of this current effort (and also employed successfully by \citealt{Burigana_et_al}).  This approach also proves convenient for studying some effects (e.g. noise correlations, bolometer time constants) which are more easily represented in the time or frequency domain.

In passing, we mention a clever alternative method proposed by \citet{2002A&A...392..369C}, where the asymmetric-beam-induced {statistical} anisotropy of the observed noisy CMB field is compared in Fourier space to statistically isotropic noise realizations to deduce the beam asymmetry, but not the complete beam {window} function.

\subsection{Rapid Monte Carlo simulation }

We designed and implemented a software pipeline to rapidly simulate planet crossing in the time domain, and then reconstruct the beam.  The \Planck\ Collaboration has implemented an extensive software infrastructure to simulate time-ordered data \citep[\textit{e.g.}][]{2006A&A...445..373R}, but these tools are, by design and optimization, intended to simulate large surveys.  Here we want to examine the beam fitting procedure by the Monte Carlo method, focusing our interest, by contrast, on the small fraction of the data  near the planets, which permits optimizations in the design of our pipeline suited to that task.
   On a laptop, our pipeline can simulate a planet crossing and subsequent beam reconstruction in a few seconds, fast enough that, on a cluster, we can rapidly generate thousands of simulations.
The modeling includes simulated pointing, realistic beams, planets, $1/f$ and white noise, the CMB, and several time-domain filters. We test our beam fitting methods using these simulated observations, characterizing the beam errors by the Monte Carlo method.  

The pointing is generated on rings and includes a randomized re-pointing error between rings.  We model nutation of the satellite spin axis as a cross-scan oscillation at a fixed frequency.   We translate the pointing into the frame where the planet is fixed, accounting for linear motions of the planet on the sky, appropriate for the few-hour time scales here.

\citet{Tauber_optics} summarizes the optical properties of the \Planck\ mission's telescope.
For the beam, we use the detailed calculations produced by the collaboration based on models of the telescope optics \citep{2002AIPC..616..242S,LFI_optics,2004SPIE.5487..542Y,HFI_optics}.
Beam values are provided on a tabulated grid, which we evaluate at non-grid points by interpolation.  We fit a Gaussian to capture the beam's largest scales, then use 2-d cubic spline interpolation to reproduce the residuals to this fit.  The interpolated beam is the sum of the Gaussian and the spline interpolation and reproduces the gridded beam exactly on pixel centers.  We model the planet as a point source, so that after convolution with the beam, the planet signal resembles the beam shape, with peak temperature given by Table~\ref{tab:planet_brightness}.

To include the impact of the CMB on the planet fits, we simulate small scale CMB modes.  These are computed by FFT in a flat sky approximation on a plane surrounding the planet scan, expanded to avoid edge effects in the beam data.  Because of the high planet signal, we find that the CMB does not have a material effect on the beam recovery.

\subsection{Detector properties}\label{sec:detprop}

Our simulated detectors are primarily characterized by their noise attributes, which we set to mimic the actual \Planck\ detectors {(see Table~\ref{tab:det})}. Optionally, for the HFI detectors, we include a time-constant and/or nonlinear response in our simulations.

To capture low-frequency drifts in the electronic amplifiers and bolometer temperatures, we use a noise power spectrum of the form
\begin{equation}
P_{\rm n}(f) = P_{\rm white} [ 1 + (f/f_{\rm knee})^{-\alpha} ]
\end{equation}
where the low frequency index $\alpha \approx 1.7$ for LFI and  $\alpha \approx 2.0$ for HFI.  We consider this noise as a sum of correlated and white parts, generated in separate steps.
The correlated low frequency part is generated via an FFT, and is continuously but slowly sampled (typically $\sim 1$ Hz) for the duration of the planet crossing (16--96 hours, depending on the beam size), then interpolated to the detector {sampling frequency} (up to $200$ Hz) {only when the detector is close to the planet}.  The white noise is sampled at the detector rate, but generated only {near the planet}.  This multi-scale approach is much faster than generating the noise at the full data rate for the duration of the crossing.  The interpolation and slow sampling of the correlated noise realization causes a smoothing of the low frequency portion of the noise, but the slow sampling rate is chosen based on the knee frequency so that the white noise masks this smoothing, and yields Gaussian noise with a very close approximation to the desired power spectrum.

In practice, \Planck's data streams will be filtered to decrease the impact of the low frequency noise, particularly in the course of mapmaking.  One promising way to achieve this is through destriping \citep[][]{1999astro.ph..6360B,1998A&AS..127..555D,1998astro.ph.10475R,2003A&A...401.1215S,2004A&A...428..287K}, which involves fitting offsets to the noise, using crossing points in the scan as points of reference to separate signal and noise.  \citet{2004MSAIS...5..419T} found for LFI detectors that undestriped $1/f$ imparts 20--30 percent systematics to the flux recovery of 1 Jy point sources (roughly Neptune's flux at 44 GHz).  The noise after destriping may be characterized with an effective power spectrum.  Using the analysis of \citet{2009A&A...506.1511K}, we express the power spectrum of this noise as
\begin{equation}
P_{\rm destripe}(f) =  
P_{\rm n}(f) \times \left[
10^{-6} + \left( 1 - \frac{\sin^2(\pi f t_{\rm off})}{(\pi f t_{\rm off})^2} \right)^2
\right]
\end{equation}
where the offset duration  parameter, $t_{\rm off}$, produces the lowest noise residuals near $1/(2 f_{\rm knee})$  (the precise minimum of the residuals depends somewhat on the details of the signal and scan).  We consider this noise as the residual $1/f$ in our planet observations.

The HFI aboard \Planck\ \citep{2003NewAR..47.1017L} consists of 52 bolometers \citep{Holmes2008} fed by feed horn structures, read out at nearly 200 Hz. The bolometer's thermal response to an the incoming optical signal is described by a transfer function expressed in the Fourier domain as a single pole low-pass filter:
\begin{equation}
T(\omega) = \frac{1}{1+i \omega \tau}
\end{equation}
where $\omega$ is the angular frequency of the signal and $\tau$ is the thermal time constant of the bolometer.  In analysis of CMB data, the detector time constant can be treated as part of the effective beam or simply deconvolved from the time ordered data as a pre-processing step \citep{1998MNRAS.299..653H}.  Here we treat the bolometer time constant as an additional parameter in our model of the instrumental response to planet observations.

In practice, the details of the bolometer's thermal circuit can lead to a transfer function that is not described by a single-pole low-pass filter.  In principle we can include a more general transfer function as additional parameters in our fit.

The ambient optical background of the HFI bolometers is dominated by the CMB and thermal emission from the telescope and optical filters.  The planets Mars, Jupiter, and Saturn are expected to be extremely bright compared to the ambient optical background and will drive the bolometers nonlinear.
In the case of HFI's readout electronics \citep{1997A&AS..126..151G}, a drop in bolometer resistance always leads to a state of ``overcompensation'' for the transient compensation.  The average signal over a readout cycle will not drop as much as in a DC biasing scheme, mitigating the effect of bolometer nonlinearity.
To simulate the nonlinear response of HFI, we use a gain curve from a  model of the detector and readout electronics \citep[DESIRE,][]{Catalano_2009}.

In practice, the response can be modeled and corrected, however when we include nonlinearity here, we will simply cut the samples most affected by nonlinearity, which should give a performance baseline that we should exceed.

\begin{table*}
  \begin{center}
    \begin{tabular}{ccccccc}\hline\hline
      Band  & Sample rate & White noise  & Knee for $1/f$  & Low-$f$  & Bolometer & FWHM \\ 
 (GHz) & (Hz)  &($\mu$K s$^{-1/2}$) & (mHz)&index $\alpha$ & $\tau$ (ms) & (arcmin.)\\
 \hline
\\
\multicolumn{6}{c}{\textit{Low-Frequency Instrument}} \\
\\
30  & 32.5  & 170   & 50 & 1.7 & \dots & 32 \\
44  & 46.5  & 200   & 50 & 1.7 & \dots & 20 \\ 
70  & 78.8  & 270   & 50 & 1.7 & \dots & 13 \\
\\
\multicolumn{6}{c}{\textit{High-Frequency Instrument}} \\
\\
100 & 185 & 50    & 30 & 2   &        10.3 & 9.2	\\
143 & 185  & 62    & 30 & 2   & 	4.5 & 6.5	\\
217 & 185  & 91    & 30 & 2   & 	3.2 & 4.5	\\
353 & 185  & 277   & 30 & 2   & 	4.2 & 4.2	\\ 
545 & 185  & 1998  & 30 & 2   &        1.5 & 4.2	\\
857 & 185  & 91000 & 30 & 2   & 	1.9	& 4.0 \\
\\
      \hline\hline
    \end{tabular}
  \end{center}
  \caption{Planck detector properties used for modeling. The HFI sampling rate is approximate and subject to on-orbit tuning. Noise figures give values in CMB thermodynamic units. FWHM based on a Gaussian fit to the realistic beam models mentioned in text. References: \citet{2006astro.ph..4069T,2009A&A...493..753A,Holmes2008}. }
\label{tab:det}
\end{table*}

\subsection{Beam model I: parametric linear distortions}\label{sec:parmodel}

In the final analysis of \Planck\ data, the observations of the planets can be used to constrain the principal components of a parameterized beam model, based on optical computation of the system of mirrors and horns.   At sufficient accuracy, this model defines a family of beams which can faithfully represent the actual beam.  Though the full analysis is beyond the scope of this work, we can proceed fruitfully using a linear approximation to the beam distortion, which can be described by seven parameters.\footnote{Very simple parametric models of the beam, like an elliptical Gaussian, cannot faithfully represent the beam simulations, and impart noticeable biases to the beam {window} function.  See Sec.~\ref{sec:tf}.}  

When tracing rays, a distortion in the optics can be represented by a transformation of the ray destinations.  A ray from the source which originally arrived at the image plane at $\mathbf{x}$ will arrive at $\mathbf{x}' = \mathbf{T}(\mathbf{x})$ after distortion.  For nearby rays and small distortions, we can expand in a series to linear order:
\begin{equation}
 \mathbf{x}' = \mathbf{T}(\mathbf{x}) = \mathbf{T}_0 + \mathbf{T}_1 \mathbf{x} + \dots
\end{equation}
where $\mathbf{T}_0$ is a vector in the plane and $\mathbf{T}_1$ is a $2 \times 2$ matrix.  This is equivalently expressed as
\begin{equation}
  \mathbf{x'} \approx \mathbf{T_1}(\mathbf{x-x_0}),
\end{equation}
where we introduce $\mathbf{x}_0$ as a beam offset {in preference to} $\mathbf{T}_0$.

Therefore, if $B_{\rm true}(\mathbf x)$ defines the realistic beam at position $\mathbf{x}$, then the distorted beam can be written as
\begin{equation}
  B_{\rm model}(\mathbf{x}) =    A \  B_{\rm true}\left( \mathbf{x'} \right)
\end{equation}
where A is the relative amplitude.
It is more convenient to work with the inverse of $\mathbf{T}_1$ and decompose its four elements into a rescaling, a rotation, and two components of shear:
\begin{eqnarray}
\mathbf{T_1}^{-1} &=&  
\left( \begin{array}{cc} 1+s & 0 \\ 0 & 1+s \\ \end{array} \right) \ 
\mathbf{R}(\psi)  \ 
\left( \begin{array}{cc} 1+\gamma_+ & \gamma_\times\\ \gamma_\times & 1-\gamma_+ \end{array}   \right)  \\ \nonumber
&& \qquad \times ({1-\gamma_+^2 - \gamma_\times^2})^{-1} 
\end{eqnarray}
where $s$ defines the rescaling, $\mathbf{R}(\psi)$ is a rotation through angle $\psi$, and $\gamma_+$ and $\gamma_\times$ are respectively the perpendicular and diagonal components of shear.  

Note that this simple parametric family contains the realistic beam, when the offset $\mathbf{x_0}$ is zero and the transformation matrix $\mathbf{T_1}$ is the identity.  Because the rotation and shear transformations have unit determinant, the $s$ parameter completely characterizes the solid angle of the beam, with $\Omega \propto (1+s)^2$.
  Because of the comparative rigidity of this model, we find that we can successfully constrain detector transfer function parameters or multiple planet amplitudes as additional parameters in a single fit.

This parameterized model, with as little as seven parameters per beam, is appealing, but probably be too simplistic.  For example, in the \citet{2009ApJS..180..246H} analysis of  \emph{WMAP}, $\sim 430$ parameters are used to fit simultaneously the ten beams of each telescope.   Our parametric model's simplicity implies a rigidity, which manifests itself in probably too optimistic beam errors.  Later we compare with a more flexible model.

\subsection{Model parameter fitting and convergence}

To fit the parameters to the data samples $d_i$, we minimize 
\begin{eqnarray}
  \chi^2& = \displaystyle \sum_{ij}& \left[ d_i - B_{\rm model}(\mathbf{x}_i) \right] C^{-1}_{ij} 
\left[ d_j - B_{\rm model}(\mathbf{x}_j) \right] \nonumber
\end{eqnarray}
(summed over time samples $i,j$) with a downhill simplex method.  There are a few $10^4$ time samples within a few beam full-width half-maximums (FWHMs) of a planet.  This makes fitting in the time domain tractable for a sparse covariance matrix, but possibly not for a dense matrix.  We only considered diagonal matrices, appropriate for white noise.  This is a simplification when  we include $1/f$ noise and CMB. In practice this does not to bias our result (averaged over an ensemble of noise and CMB realizations), although it makes the errors on the fit larger than with an optimal estimate.

In our parametric beam model, the vector of parameters which exactly recovers the true beam is 
\begin{equation} \{A, \mathbf{x}_0, s, \psi, \gamma_+, \gamma_\times\} = \{ A_{\rm planet}, (0',0'), 0, 0^\circ, 0, 0\}. \end{equation}
Because of noise and sparse pointing, the set of parameters which minimizes $\chi^2$ will differ from these, and the distribution of these parameters characterizes our uncertainty in the beam recovery.  The sources of noise (detector and CMB) are Gaussian, so the $\chi^2$ minimization will seek an unbiased estimate for the beam function at each sample.  The beam parameters are related to the beam via a non-linear function, so they can be biased in our estimate, even when the beam is unbiased.  In practice, however, we find these biases to be negligible.  For example, fitting to Jupiter after $\sim 10^3$ MC steps for our highest planet-signal-to-noise channel (857 GHz), the mean cross-scan position is about 1.5 times the predicted standard deviation of the mean for this sample size.  This bias is 5 percent of the typical dispersion for a single realization, and only $10^{-7}$ times the beam FWHM.  Other channels show similar biases, and the biases on other parameters are similar or better.

At each step in the Monte Carlo simulation (each with an independent realization of pointing, noise, and CMB), we start the minimization at a random point in parameter space, located with uniform probability within a rectangular solid centered on the parameters of the true beam, with dimensions $\{ 0.5 \times A_{\rm planet}, (2',2'), 0.2, 10^\circ, 0.2, 0.2\}$.  The dimensions of this box, cut in half, are used for the step size to initialize the simplex minimization\footnote{We use an implementation (\texttt{nmsimplex}) from the GNU Scientific Library, version 1.10}.  We execute the algorithm for a fixed number of iterations, which sets the accuracy of the minimization at each step in the Monte Carlo.  Parameter distribution convergence depends on this accuracy and on the total number of Monte Carlo steps.

To verify the convergence of our beam parameter distributions, we ran a suite of simulations, varying the iteration count in the minimization.  We consider one horn from every band of \Planck\ observing Jupiter, with destriped noise and CMB, including a time constant for the HFI detectors.  Depending on the channel, we find that after 1,000--3,000 iterations, the simplex minimization will have settled sufficiently to not have a material effect on the variance of the parameter distributions.  (More parameters require more iterations to converge.)  The 353 GHz beam is the slowest to converge in the simplex fitting, and the only channel which occasionally settles in spurious local minima in $\chi^2$ (more fully discussed in Sec.~\ref{sec:parresults}).  The variance of beam parameters converges after roughly 1000 Monte Carlo steps.
%
%
%
%

\subsection{Beam model II: non-parametric basis functions}

We have argued that our parametric model may be too rigid to give realistic beam errors.  As an alternative, we want to constrain the beam directly by the planet measurements, without recourse to an optical model.  Parametric methods are more powerful statistically, but non-parametric reconstruction should provide a robust consistency check.  {In addition, if the beams contain tails, corrugations, or other features  on-orbit which were not present in ground measurements or simulations, the non-parametric method provides a way to represent them.}

{To represent an arbitrary beam, any} complete basis will suffice, but some represent the beam more compactly.  The {simulated} beams of most channels are approximate elliptical Gaussians, so the eigenfunctions for an asymmetric 2-D quantum harmonic oscillator form a convenient basis.  We construct the ground state of the Hermite-Gauss functions, an elliptical Gaussian, so that it minimizes the square deviation from the simulated beam.  Below we illustrate the computation of basis coefficients with integration on the sparsely sampled sky using the orthogonality relation, as opposed to fitting for parameters.

\begin{figure*}
\begin{center}
\includegraphics[width=0.45\figwidth]{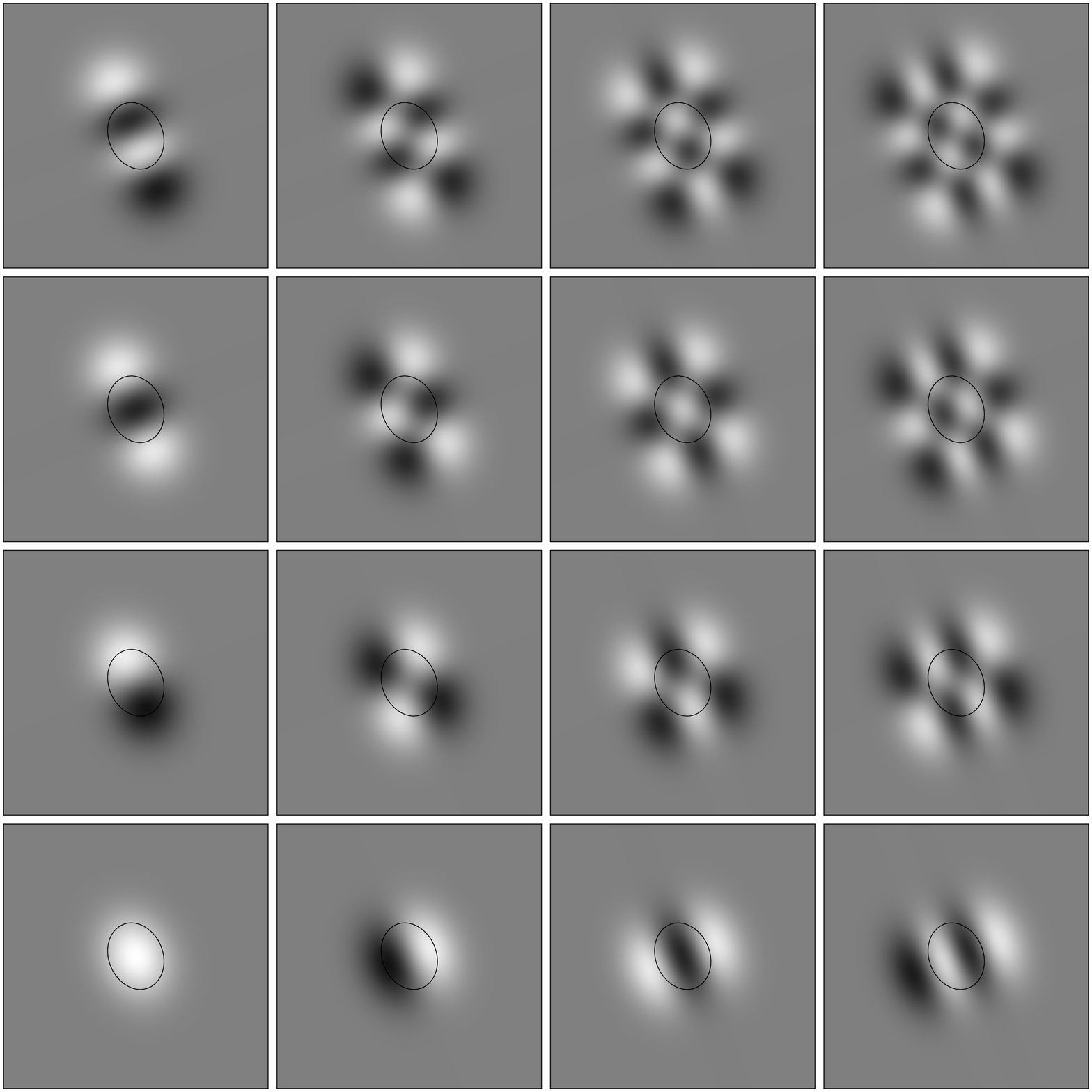}
\includegraphics[width=0.5\figwidth]{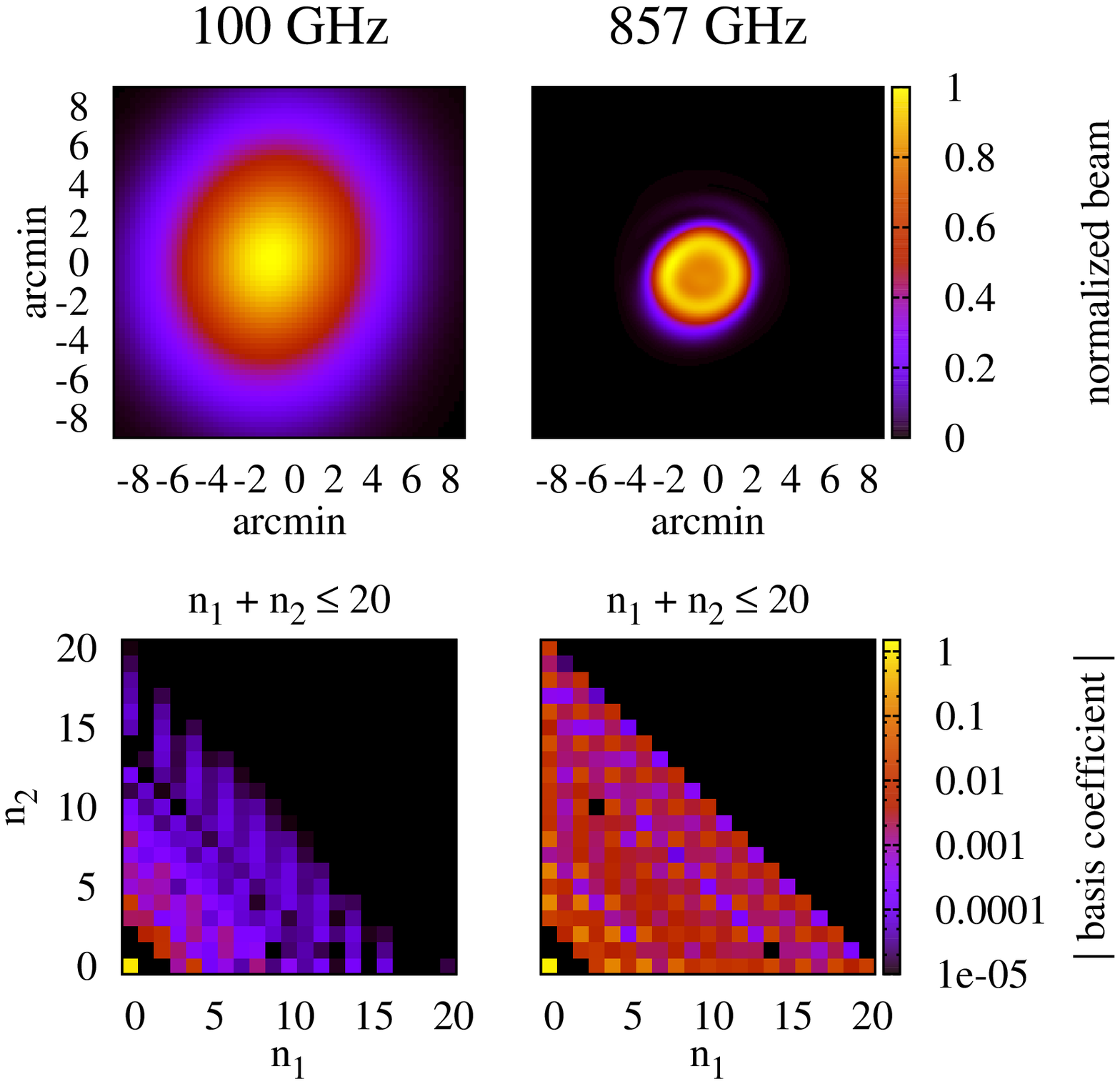}
\end{center}
\caption{Basis suitable for decomposing elliptical beams.  At left: the orthogonal and complete basis functions, where the ellipse marks the half-maximum curve Gaussian portion of the basis (bottom left).  Every subpanel to the right or up increments the index on the Hermite polynomial perpendicular to or along the ellipse's major axis. At right: efficiently and inefficiently represented sample \Planck\ beams, and their decomposition into the basis, limited to $n_1+n_2 \leq 20$.  
}\label{fig:beamshapelets} \label{fig:beam_coef}
\end{figure*}

Similar basis functions (but based on an axisymmetric Gaussian) are frequently used in astronomy, for example to describe galaxies shapes \citep[``shapelets,''][]{2003MNRAS.338...35R,2005MNRAS.363..197M} and telescope PSFs in gravitational lensing studies  \citep{2002AJ....123..583B,2003MNRAS.343..459H}.
In particular, \citet{2003MNRAS.338...35R} considers distortions of the axisymmetric basis which is essentially  our approach here, in a different context. 
A sample set of beam basis functions are shown in Fig.~\ref{fig:beamshapelets}.

We argue that this basis has several advantages for representing beams, and in particular we contrast it with Zernike polynomials, which are commonly used as a basis set to represent distortions on the surface of an optical element, and are sometimes used to represent beams.  Elliptical Hermite-Gauss functions, by design, reproduce an elliptical Gaussian beam with the first basis coefficient, compared to Zernike polynomials, which take many more components to approximate it.  Elliptical Gaussian beams are a frequently used and well-studied approximation for off-axis CMB instruments.  Hermite-Gauss functions are orthogonal over the whole plane while Zernike polynomials are limited to a disk, the reason they are convenient for distortions on circular apertures.  This is not  a problem for the Zernike basis, if we scale the disk larger than the area where we consider beam data, but the size of the disk we should use is ambiguous, particularly for beams with a wide variety of sizes.  This same scale ambiguity is constrained in the our elliptic Hermite-Gauss basis by fitting the ground state to the beam.

We checked that most of the simulated \Planck\ beams, and the ones most important for CMB measurement, may be represented compactly in the elliptical Hermite-Gaussian basis (again, see Fig.~\ref{fig:beam_coef}).  However, the multimoded horns (at 545 and especially 857 GHz) do not closely resemble elliptical Gaussians, and this basis is less effective when a small number of basis functions are employed.  {Because the basis is complete, these beam can be represented by pushing to higher order, but it may be inefficient and unappealing to do so.}
{Possibly}, a different basis may be more successfully applied to these channels, but we have not explored this.  {For the channels where this basis does work well, the beam data should always be tested with higher order modes than the optics simulations require, checking for evidence of unexpected features at very low signal.}

Elliptical Hermite-Gauss basis functions are written as
\begin{equation}
\Phi_{n_1 n_2}(\mathbf{x}) = \frac{H_{n_1}(x'_1) H_{n_2}(x'_2)}{ \sqrt{2^{n_1+n_2} n_1! n_2!}} \exp \left( - \mathbf{x' \cdot x'}/2 \right)  
\end{equation}
where $H_n$ is the Hermite polynomial of order $n$.  The parameters of the best-fit elliptical Gaussian are encoded in the transformation
\begin{equation}
\mathbf{x'} =
\left( \begin{array}{cc} \sigma_x^{-1} & 0 \\ 0 &\sigma_y^{-1} \end{array} \right)
\mathbf{R}^{-1}(\psi) 
\left( \mathbf{x-x_0} \right), 
\end{equation}
which offsets the beam position by $\mathbf x_0$, rotates the beam through $\psi$, and scales the ellipse axes by 
\begin{eqnarray}
\sigma_x =& t^{-1/2} &\times \ \  \theta_{\mathrm{FWHM}} / \sqrt{8 \log 2} \\ \nonumber
\sigma_y =& t^{1/2} &\times \ \ \theta_{\mathrm{FWHM}} / \sqrt{8 \log 2},
\end{eqnarray}
where $t$ is elliptical Gaussian's axis ratio and $\theta_{\mathrm{FWHM}}$ is the geometric mean full width at half maximum.
Then we can expand the beam as
\begin{equation}
B(\mathbf{x}) =  \sum_n s_n \Phi_n (\mathbf{x}),
\end{equation}
where we re-index the eigenmodes with 
\begin{equation}
n = \frac{(n_1+n_2)^2 + n_1 + 3n_2}{2}.
\end{equation}

With the normalization used here, convenient for beams, the maximum of the ground state function is unity, and the orthogonality relation is
\begin{equation}
\int d^2x \ \Phi_{m} (\mathbf x) \Phi_{n}  (\mathbf x) = \frac{\pi  \theta_{\mathrm{FWHM}}^2 \delta_{m n}}{8 \log 2}.
\end{equation}
So  to recover the basis coefficients $s_n$, one integrates
\begin{equation}
s_n = \frac{{ 8 \log 2}}{ \pi \theta_{\mathrm{FWHM}}^2} \int d^2x \ \Phi_{n}  (\mathbf x) B  (\mathbf x).
\end{equation}

Since \Planck\ only samples the sky (Fig.~\ref{fig:pointing}), these integrals must be approximated as sums over samples
\begin{equation}
 \int d^2x \ (\dots) \approx  \frac{A}{N} \sum_i \ (\dots)
\end{equation}
computed from the $N$ detector samples from the area $A$ surrounding the planet.  In this sampling, the modes are only approximately orthogonal.  The approximation becomes poor if the typical size of the gap in the sampling is large compared to the FWHM of the Gaussian on which the functions are based, or compared to the scale of oscillations in the highest frequency modes considered.  Uncorrected, this will lead to biased estimates for the beam coefficients.

We address this by computing, to some maximum relevant mode, the symmetric overlap matrix of the basis functions on the sampled sky,
\begin{equation}
I_{mn} = \frac{{ 8 \log 2}}{ \pi \theta_{\mathrm{FWHM}}^2} \frac{A}{N} \sum_i  \ \Phi_{m} ({\mathbf x}_i) \Phi_{n} ({\mathbf x}_i).
\end{equation} 
For the plane subsampling, beams, and maximum mode for a typical \Planck\ beam, this is a dense matrix, with diagonal entries with values near 1 and off-diagonal entries ranging from roughly $-0.2$ to $+0.2$.
To get unbiased beam coefficients, we use the inverse of this overlap matrix to deconvolve, yielding an estimate of the basis coefficient:
\begin{equation}
  \hat s_m = \sum_n \ I^{-1}_{mn} \frac{{ 8 \log 2}}{ \pi \theta_{\mathrm{FWHM}}^2} \frac{A}{N} \sum_i   \Phi_{n}  (\mathbf{x}_i) d_i,  \label{eqn:estimate_basis_coef}
\end{equation}
where the $d_i$ are the time-ordered data, consisting of signal and noise,
\begin{equation}
  d_i = \sum_n s_n \Phi_n(\mathbf{x}_i) + n_i.
\end{equation}
In the limit of a well sampled beam, the sums closely approximate the continuous integrals, and $I_{mn}$ approaches the Kronecker $\delta_{mn}$. Adding zero mean noise does not bias the estimated basis coefficients in the ensemble average.  If the noise is white, characterized by a covariance $ {\rm Cov}(n_i, n_j) = \sigma^2 \delta_{ij}$, then the covariance of the estimated basis coefficients is 
\begin{equation}
{\rm Cov}(\hat s_m, \hat s_n) = \frac{{ 8 \log 2}}{ \pi \theta_{\mathrm{FWHM}}^2} \frac{A}{N} \sigma^2 I^{-1}_{mn}.
\end{equation}

\begin{figure}
\begin{center}
\includegraphics[width=\figwidth]{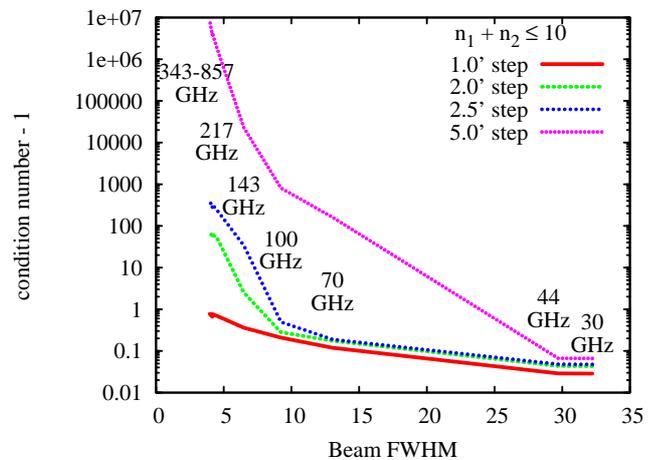}
\end{center}
\caption{ For varying re-pointing steps, the overlap matrix $I$ condition number for modes with $n_1+n_2 \leq 10$, for the Planck beams. \label{fig:condnum} }
\end{figure}

In the linear operation which corrects for the sampling (represented by $I^{-1}$ in Eq. \ref{eqn:estimate_basis_coef}), the condition number of the overlap matrix quantifies how well a particular sky sampling supports a set of basis functions.  
For a range of beam sizes and pointing realizations, we have plotted the overlap matrix condition number in Fig.~\ref{fig:condnum}.
Sparser sampling, for a fixed overall number of sampling points, boosts the noise in the beam coefficient reconstruction proportional to the condition number.
Thus study of the overlap integral provides a means to evaluate quantitatively the effect of a given scanning strategy on the quality of the beam reconstruction.

\subsection{Beam window functions} \label{sec:tf}

For both our reconstruction techniques, our understanding of the beam impacts the cosmological interpretation of the CMB power spectrum:  we deconvolve the beam to get an unbiased estimate.  Define the beam {window} function as the function $b^2_l$ such that the ensemble average of the measured power spectrum relates to the true power spectrum as $  \langle \tilde C_l \rangle = b^2_l C_l $.  To make an unbiased estimate of the power spectrum, we divide the observed spectrum by $b^2_l$ at each multipole.

For an asymmetric beam and an arbitrary scan strategy, computing the {window} function efficiently remains an open research question \cite[see e.g.][]{2007ApJS..170..288H,2008PhRvD..77h3003S}.  However, we can make progress with suitable approximations. Over the course of a single survey, the \Planck\ scanning strategy is approximately pole-to-pole, so that over most of the sky (except near the poles) the beam strikes at a common orientation.  The single orientation {window} function is readily computed in the flat sky approximation, where we can express the beam convolution as a multiplication in Fourier space.  With 2-d wavenumber $\mathbf l$, the measured signal is $ B(\mathbf l) s(\mathbf l)$ and we write the measured power spectrum as an average over azimuthal angle on the Fourier plane:
\begin{equation}
\tilde C_l = \frac{1}{2\pi} \int_0^{2\pi} dl_\phi \   B(\mathbf l) s(\mathbf l) B^*(\mathbf l) s^*(\mathbf l) 
\end{equation}
Then the ensemble average yields the {window} function,
\begin{eqnarray}
\langle \tilde C_l \rangle&=& \frac{1}{2\pi} \int_0^{2\pi} dl_\phi \   B(\mathbf l)  B^*(\mathbf l) \langle  s(\mathbf l) s^*(\mathbf l) \rangle \nonumber \\ \nonumber
&=&  \left[ \frac{1}{2\pi} \int_0^{2\pi} dl_\phi \   B(\mathbf l)  B^*(\mathbf l) \right] C_l  =  b^2_l C_l,
\end{eqnarray}
where the term in brackets gives {window} function.  We enforce the normalization $b_l \rightarrow 1$ as $l \rightarrow 0$.  (In practice we normalize the lowest $l$-bin.)  Compared to the true scanning strategy for \Planck, this type of approximation imposes an error of fraction of a percent onto the temperature {window} function \citep{2009A&A...493..753A}.
If we include a detector transfer function in our fitting, we additionally convolve the beam map with a spatial filter representing the effect of the detector's response during a constant velocity scan, then deconvolve a filter based on the (slightly-different) fitted transfer function.

\begin{figure}
\includegraphics[width=\figwidth]{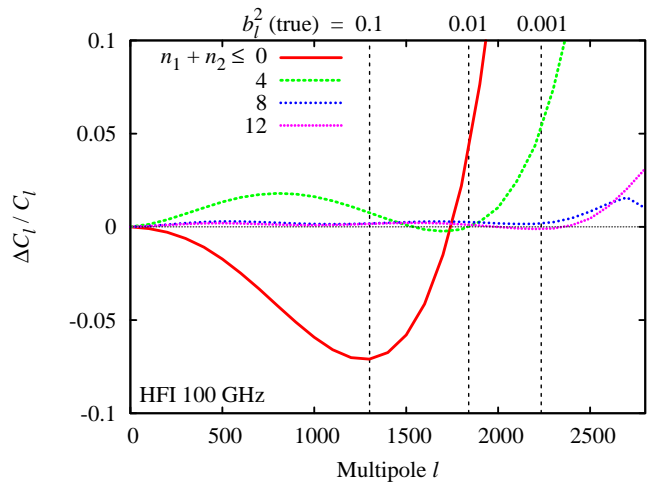}
\caption{Bias in the power spectrum caused by limiting the number of basis functions for beam reconstruction, showing the poor result from the elliptical Gaussian beam  approximation ($n_1 = n_2 = 0$).  For HFI 100 GHz.} \label{fig:transfunc}
\end{figure}

The mistake we make when we deconvolve an approximate beam is the ratio of the reconstructed {window} function to the true {window} function: \begin{equation}r^2_l =  b^2_l /  b^2_{l,\rm{true}}.\label{eqn:transrat} \end{equation}
The fractional error in the power spectrum is then \begin{equation}\Delta C_l/C_l = r^2_l - 1.\end{equation}
Fig.~\ref{fig:transfunc} show this quantity for limited numbers of eigenfunctions of the non-parametric models in one of the 100 GHz channels (assuming no noise and dense sampling of the beam).  Elliptical Gaussians fit to the simulated beam in real space make a poor approximation for computing the {window} function.

For a set of Monte Carlo realizations of the fitted beam, we can compute the set of {window} function ratios, and characterize the covariance matrix due to beam reconstruction errors\footnote{One peculiarity of our approximation is that only three parameters of the parametric model contribute to the {window} function.  It depends only on the scale (beam solid angle) and two components of shear.  The other four parameters in the model do not contribute: the amplitude is an overall calibration; the position offsets translate to phases in Fourier space, and cancel in the multiplication of complex conjugates; and the rotation is integrated out.  Indeed the beam covariance matrix in this case has only three substantial eigenvalues.}:
\begin{eqnarray}
{\rm Cov}(\tilde C_{l_0},\tilde C_{l_1}) &=& C_{l_0} C_{l_1} \left\langle ( r^2_{l_0} - 1)(r^2_{l_1} - 1 ) \right\rangle \\
&  \approx & \frac{C_{l_0} C_{l_1}}{N_{\rm MC}} \sum_i ( r^2_{l_0,i} - 1)(r^2_{l_1,i} - 1 ).  \nonumber
\end{eqnarray}
The covariance matrix is dense, strongest on the diagonal and smoothly dropping away from it.  Beam errors are highly correlated across multipoles.  We find the correlation coefficients by normalizing the diagonal of the covariance matrix.  For the parametric model at 100 GHz, the correlation coefficient between high ($l \sim 3000$) and low ($l\sim100$) multipoles is very strong, at least 0.88.  This is important for the measurement of cosmological parameters, such as $n_s$ or $\tau$, which couple across large multipole ranges.

Some caution is appropriate regarding bias in the recovery of the {window} function.  The {window} function is a quadratic function of the beam.  Therefore a procedure which reconstructs an unbiased estimate of the beam can produce a biased estimate of the {window} function.  For our beam recovery techniques, we in practice detect little bias in the ensemble of {window} functions for channels at 100 GHz and below.  In higher bands, with smaller beams and higher signal-to-noise, the mean of the Monte Carlo ensemble (per $l$) slowly oscillates near, but slightly above or below the true {window} function.  In principle a correction to this bias can be folded into the power spectrum analysis, but here we allow the bias to persist, to see if it has any effect in the cosmological analysis.

\section{Results} \label{sec:results}

Here we compile results for beam fitting and the subsequent errors imposed onto the CMB power spectrum.  
We take a single Jupiter crossing as our baseline case for beam fitting, and consider a case  with destriped $1/f$ noise, with no contribution from large-scale ($l<250$) CMB.  (Confusion from signals on the sky can be removed because every region of the sky is re-observed at 7-month intervals).  For each frequency band, we use one model LFI or HFI beam.  Unless noted, we assume nonlinearities in the detector response have been corrected before processing.

\subsection{Parametric model}  \label{sec:parresults}
\begin{table*}
\begin{center}
\begin{tabular}{ccccccccc} \hline\hline
Band  & $\sigma_{\rm cross}$ & $\sigma_{\rm co}$ & $\sigma_{\rm FWHM}/{\rm FWHM}$  &$\sigma_\psi$  & $\sigma_{\gamma_+}$  & $\sigma_{\gamma_\times}$ & $\sigma_A/A$ & $\sigma_\tau/\tau$\\
(GHz)  &  \multicolumn{2}{c}{($10^{-4}$ arcmin.)} & ($10^{-6}$) & ($10^{-2}$ deg.) & ($10^{-5}$) &  ($10^{-5}$) &  ($10^{-5}$) & ($10^{-5}$) \\ \hline
\\
\multicolumn{9}{c}{\textit{Low-Frequency Instrument}} \\
\\

30  & $218$  & $344$    & $937$ & $267$    & $548$   & $1270$  & $164$   & \dots \\
44  & $125$  & $152$    & $572$ & $53.1$   & $143$   & $118$   & $92.1$  & \dots \\
70  & $26.6$ & $35.6$   & $418$ & $34.3$   & $38.9$  & $141$   & $51.6$  & \dots \\
\\
\multicolumn{9}{c}{\textit{High-Frequency Instrument} (known $\tau$)} \\
\\
100 & $7.24$  & $25.3$  & $157$ & $13.0$   & $19.2$  & $36.1$  & $10.1$  & $0.00$   \\
143 & $3.74$  & $6.76$  & $69.7$ & $4.30$  & $4.68$  & $6.17$  & $4.77$  & $0.00$  \\
217 & $0.473$ & $2.26$  & $23.4$ & $3.86$  & $9.96$  & $3.86$  & $2.26$  & $0.00$ \\
353 & $8.96$  & $9.81$  & $13.3$ & $4.72$  & $22.9$  & $2.86$  & $1.83$  & $0.00$  \\
545 & $0.295$ & $0.167$ & $5.15$ & $0.167$ & $0.763$ & $0.521$ & $0.866$ & $0.00$ \\
857 & $0.351$ & $0.153$ & $1.87$ & $0.287$ & $0.454$ & $0.239$ & $0.398$ & $0.00$  \\
\\
\multicolumn{9}{c}{\textit{High-Frequency Instrument} (fitting $\tau$) } \\
\\
100 & $7.13$  & $37.3$  & $144$ & $13.8$   & $18.7$  & $37.7$  & $33.0$  & $96.3$ \\
143 & $4.41$  & $15.6$  & $80.8$ & $5.36$  & $6.91$  & $7.10$  & $18.3$  & $88.8$ \\
217 & $0.494$ & $4.11$  & $30.0$ & $5.03$  & $12.3$  & $4.75$  & $6.18$  & $26.7$ \\
353 & $49.3$  & $52.0$  & $72.2$ & $25.8$  & $126$   & $20.7$  & $13.7$  & $34.1$ \\
545 & $0.300$ & $0.341$ & $5.49$ & $0.172$ & $0.828$ & $0.521$ & $0.970$ & $6.25$ \\
857 & $0.351$ & $0.164$ & $1.87$ & $0.287$ & $0.454$ & $0.238$ & $0.412$ & $1.38$ \\
\\\hline\hline
\end{tabular}
\end{center}
\caption{Standard deviations on parameter estimates from a single Jupiter observation.  The beam offset error in the cross-scan direction is $\sigma_{\rm cross}$, and similarly for $\sigma_{\rm co}$.  The beam angle $\psi$, shears $\gamma_+$ and $\gamma_\times$, and amplitude $A$ are defined in Sec.~\ref{sec:parmodel}.  The fractional FWHM error is derived from the error on the scale parameter $s$, also defined there.  The time constant $\tau$ is defined in Sec.~\ref{sec:detprop}.}
\label{tab:vanillaparameters}
\end{table*}
The  errors on the recovered model parameters are shown in Table \ref{tab:vanillaparameters} {for 7- and 8-parameter beam models, based on 1280 Monte Carlo steps with 2000 fitting iterations each (3000 in case of 353 GHz)}.  The quality of the parameter recovery is exceptionally good, due to the very high signal-to-noise on Jupiter.  Comparing channels, the relative quality of the fits is a complicated interaction between Jupiter's signal, the detector's noise, and the beam size (small beams concentrate the signal but yield fewer useful data points).  Except for 353 GHz, the quality of the fits tend to improve with increasing frequency band, largely due to the increase in Jupiter's signal at high frequencies.  Repeated observations of the planets during subsequent surveys reduce the errors roughly as expected for independent observations.

The errors at 353 GHz, although quite small, are puzzlingly larger than the errors at 217 GHz and 545 GHz, especially when fitting for a time constant.  This seems to be due to the $\chi^2$ minimization getting caught in local
 minima away from the global minimum, which creates a population of
 outliers in the Monte Carlo ensemble of fitted parameters, driving up the errors. 
 \begin{figure}
\includegraphics[width=\figwidth]{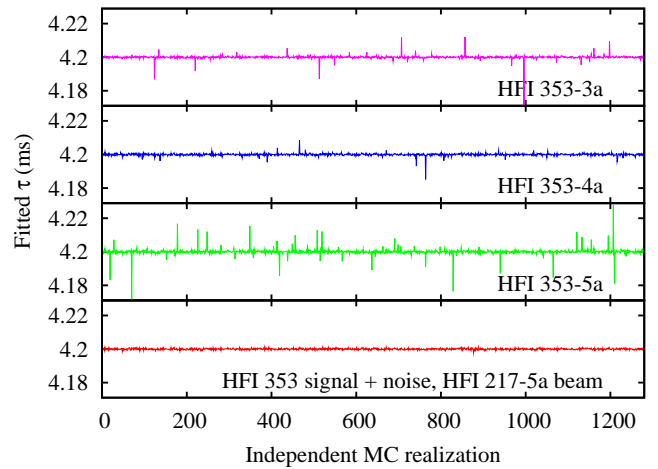}
 \caption{Parameter fits for the detector time constant for the HFI 353 GHz channel.  Several 353 GHz horns show outliers (top three panels) not seen in the other channels.  The outliers are not seen when a simulated beam from the 217 GHz channel is substituted into the 353 GHz simulation (bottom panel).}
\label{fig:whats_with_353}
\end{figure}
 These outliers represent $\sim 5\%$ of samples and are seen only in the simulated 353 GHz beams.  Although the other beams
 span a large range of beam sizes, signal-to-noise, and time constant duration, none show any obvious population of outliers.
As noted in Fig.~\ref{fig:whats_with_353}, several 353 GHz beams show this behavior.  
The outliers do not fall into any well-separated population which
 make them easy to cut, and we have not found a way to eliminate them
 robustly. 

 The 353 GHz case (signal, noise, and time constant) run with the similarly-sized 217 GHz beam
  \emph{does not} show a large population of outliers.  Visual inspection yields nothing obviously wrong with the 353 GHz beams, which are formatted the same way as the other HFI beams, and the timelines the 353 GHz beams produce in our pipeline are also unremarkable, so it remains unclear why these outliers occur.
  Even with the outliers, the beam fits for 353 are still quite
 good, only suffering by comparison to the spectacular results from
 the neighboring channels.

\begin{figure}
\includegraphics[width=\figwidth]{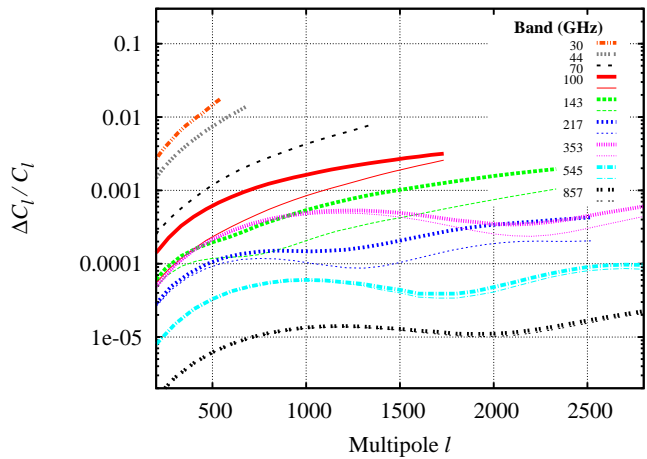}
\caption{Errors on the window function using the parametric beam model.  At each multipole, 68\% of the fitted Monte Carlo {window} functions recover spectra closer to the true power spectrum than the indicated line.  Lines are cut off where the {window} function falls to 1\%.  For the HFI (bolometer) channels thinner lines of the same color and type denote fixing the time constant before fitting, showing smaller errors.}
\label{fig:tf-par}
\end{figure}

For all channels, the corresponding errors on the power spectrum due to the {window} function uncertainty are depicted in Fig.~\ref{fig:tf-par}.  Our ensemble of {window} functions presents a slightly biased estimate of the true {window} function (section~\ref{sec:tf}), so instead of a standard deviation, we plot a contour which bounds the error for 68\% of the {window} functions in our ensemble, including both bias and dispersion.  Errors are strongly correlated between multipoles.  Depending on multipole, prior knowledge of the detector time constant improves the errors on HFI channels 100--353 GHz by a factor up to 2--3.  The errors on the higher frequency channels are less affected.  In general, the recovery of the {window} function for the higher frequency bands is exquisite, but this is a result of the rigidity of the model, due to the small number of parameters.

\subsection{Non-parametric decomposition}
\begin{figure}
\includegraphics[width=\figwidth]{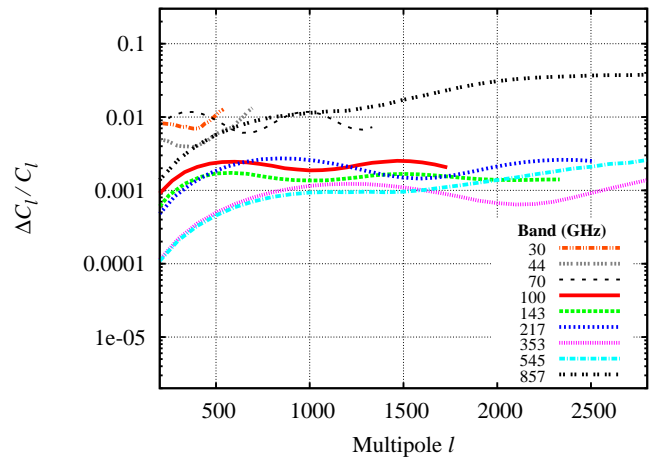}
\caption{Errors on the window function, like Fig.~\ref{fig:tf-par}, but using the non-parametric beam model limited to $n_1 + n_2 \leq 20$.}
\label{fig:tf-nonpar}
\end{figure}

The non-parametric model (Fig.~\ref{fig:tf-nonpar}) has notably larger errors.  This is due to the flexibility of this model compared to the parametric one.  For computational efficiency, we limit {the basis coefficients to those with} $n_1 + n_2 \leq 20$.  At that refinement, the 857 GHz channel in particular is poorly resolved by this basis, leading to large errors in the {window} function.

If the parametric model, where only a handful of numbers can describe the beam, describes our best case realistic scenario, the non-parametric model, requiring no prior knowledge of the beam, represents our {most conservative} case.  For the bands 100-217 GHz, which are the most important in terms of raw sensitivity to the CMB, one of the twice-annual crossings of Jupiter should yield a measurement of the beam {window} function {to within} 0.3\%.  This compares to roughly 0.5\% from 5 years of Jupiter mapping from WMAP \citep{2009ApJS..180..246H}.  Over the mission lifetime, Jupiter will be visible about four times.  Errors will decrease somewhat faster than the square root of the number of observations in the non-parametric model, because filling in the plane constrains which modes can contribute, improving the overlap matrix.

Because we have some knowledge of the beams from ground tests and numerical models, these parametric and non-parametric cases should bracket the range of reasonable possibilities.  

\subsection{Detector non-linearity}
For the parametric model, we evaluate an extremely conservative method for dealing with the nonlinear gain of the HFI, excluding data where the gain deviates from linearity by more than the rms noise per sample.  In practice, the HFI analysis pipeline will correct the data for the nonlinear response.

\begin{figure}
\includegraphics[width=\figwidth]{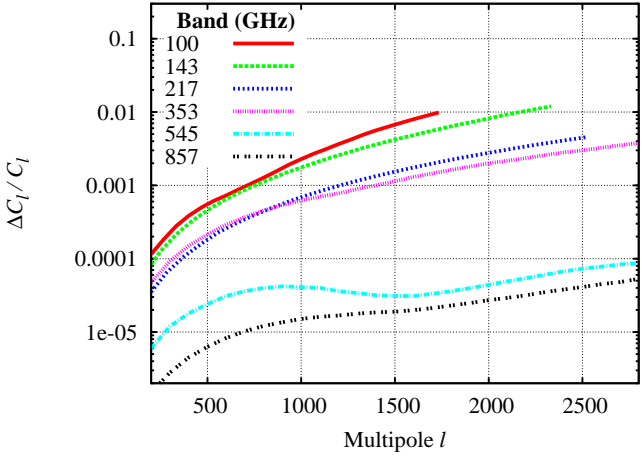}
\caption{Errors on the window function, like Fig.~\ref{fig:tf-par}, using the parametric beam model, but including the nonlinear gain of HFI and multiple planets.
}
\label{fig:tf-nonlinear}
\end{figure}

For most channels, this cut removes the peak region of Jupiter and Saturn, but keeps the central region of Mars, which provides information on the beam's peak, so we include all three in the fit.  Separate amplitudes are fit for each of the planets, and are effectively marginalized out in the computation of the {window} function.  Only the 353 GHz channel has significant nonlinear response at the peak of Mars; for this channel only we additionally include observations of Uranus and Neptune to aid the fitting.  The corresponding error in the {window} function is shown in figure \ref{fig:tf-nonlinear}, {based on 384 Monte Carlo simulations per channel}.

Although we are including more planets, the exclusion of the peak of the beam on Jupiter, where signal-to-noise is highest, boosts the noise significantly.  In this case the error on the {window} function on the small scale end of the beam is raised by a factor of 2--25, with 545 GHz the least and 217 GHz the most affected.  Correcting for the non-linearity will allow much better performance.

\subsection{Noise and destriping}

Destriping to reduce noise is an important step in beam reconstruction.
We ran cases with unfiltered low frequency noise, which manifests as stripes across the face of the beam, and this will increase the noise in the beam fit or decomposition. The non-parametric model is somewhat more sensitive to low-frequency noise than the parametric model, because the correlated noise will be mistaken for the true structure of the beam.  
 Larger beams are also more affected than small beams, because they take longer to transit the planet.  For both beam reconstruction models, low frequency noise increases the {window} function errors at 30 GHz (FWHM $32'$) by nearly two orders of magnitude and at 217 GHz (FWHM $6.5'$) by less than an order of magnitude.  The parametric model errors increase at 857 GHz (FWHM $4'$) by several tens of percent, but the non-parametric model's errors at 857 GHz are driven by the poor decomposition into basis coefficients, and are not much affected by the destriping.

\section{Implications for cosmology} \label{sec:cosmology}

We explore the tilt of the scalar perturbation spectrum with the likelihood in a simplified case where the likelihood function is analytic.  We take slices through the likelihood, modified by the beam errors, and leave a fuller Markov Chain Monte Carlo {evaluation} of the likelihood to other work \citep[see][]{2010A&A...513A..23R}.

\begin{figure}
\includegraphics[width=\figwidth]{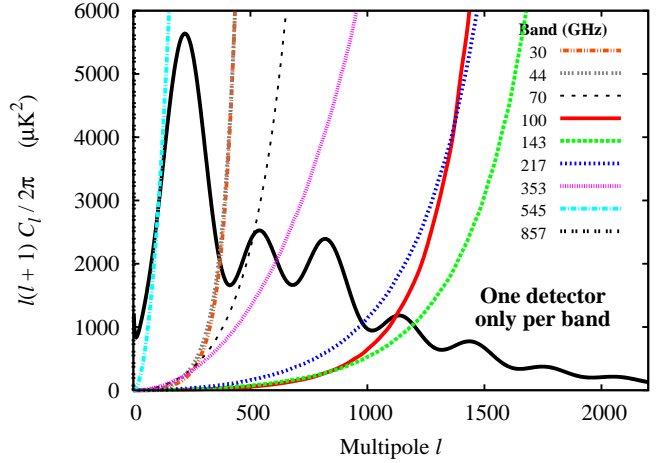}
\caption{Noise power spectra compared to the CMB temperature power spectrum, for single horns in each of the Planck channels, assuming uniform white noise (from Table \ref{tab:det}) and 1 year survey duration.  (Noise figures for the full Planck focal plane, with 74 detectors, will be much lower.)} \label{fig:detnoise}
\end{figure}

For a theoretical power spectrum, ${\cal C}_l = l(l+1)C_l/2\pi$, given a beam deconvolved data spectrum ${\cal D}_l$, which includes isotropic noise with power spectrum ${\cal N}_l$, the full-sky likelihood is \citep[e.g.][]{2000ApJ...533...19B}
\begin{eqnarray}
-2 \log {\cal L}( {\cal D}_l | {\cal C}_l ) = \\ \nonumber 
\sum_l (2l+1) \bigg[& \log ({\cal C}_l + {\cal N}_l)  +  {\cal D}_l  / ({\cal C}_l + {\cal N}_l) \bigg].
\end{eqnarray}
We assume a flat prior on ${\cal C}_l$, so that the posterior probability is proportional to the likelihood: $P({\cal C}_l |  {\cal D}_l ) \propto {\cal L}( {\cal D}_l | {\cal C}_l )$.
For the beam deconvolved noise spectrum, we take
\begin{equation}
  {\cal N}_l = \frac{l(l+1)}{2\pi b^2_l} \frac{4 \pi \sigma_t^2}{T_{\mbox{surv}}},
\end{equation}
which is exact for uniform white noise, and depends only on the time domain noise variance ($\sigma_t^2$, see Table \ref{tab:det}) and the duration of the survey ($T_{\mbox{surv}}$) which we take as one year.  To check the likelihood for individual detectors, we construct noise power spectra for one detector at a time (Figure~\ref{fig:detnoise}).  

\subsection{Gaussian beam model}

To gain intuition on the magnitude of the beam impact on parameters, we use a {1-parameter} symmetric Gaussian model for the beam {window} function, $b^2_l = \exp(-\mbox{constant} \times l^2)$, where the constant describes the width of the beam.  Deconvolving a mismatched beam yields a ratio of {window} functions which may be parameterized by the fractional error in the FWHM, which we denote $\Delta$.  For small errors, the {window} function ratio in this case is
\begin{equation}
  r^2_l = \exp\left[ 2\Delta \log a \times \frac{l^2}{ l_{a}^2} \right]
\end{equation}
which corresponds to $r^2_{l_a} = 1 + 2\Delta \log a$ at the multipole defined by $b^2_{l_{a}} = a$, where the beam {window} function has fallen by a factor $a$.
We include data only for $l<l_{0.01}$, that is, scales larger than where the beam {window} function has fallen to 1\%.

\begin{figure}
\includegraphics[width=\figwidth]{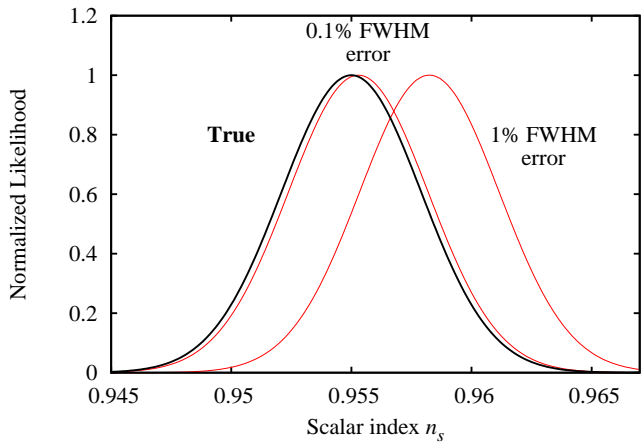}
\caption{Distortions to the likelihood slice for the scalar index $n_s$ when the fit for the FWHM of a Gaussian beam is too small by 0.1 and 1 percent, for HFI 100 GHz noise, with all other cosmological parameters fixed.  } \label{fig:simplelikeslice}
\end{figure}

\begin{figure}
\includegraphics[width=\figwidth]{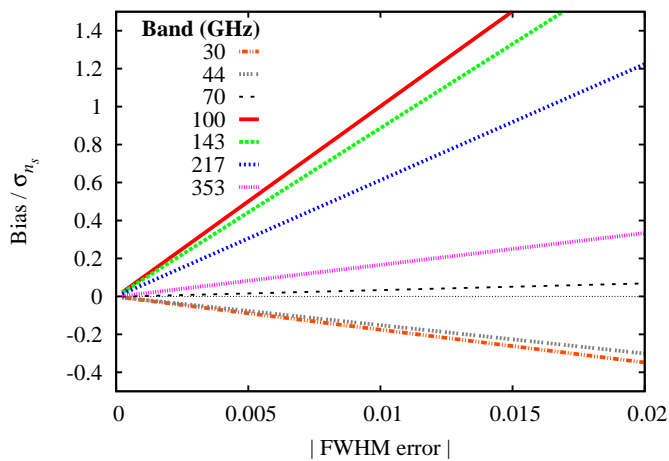}
\caption{Bias in the peak of the likelihood slice for $n_s$ in units of the error, for single horns in a Gaussian beam model, when the fitted FWHM is too small.  Horns at 545 and 857 GHz are not shown because the noise is too large to effectively measure $n_s$.} \label{fig:simplelikebias}
\end{figure}

We fix all parameters except $n_s$, which we expect to be the most sensitive to errors in the beam.  For an incorrect {window} function, the data {and  the likelihood are distorted}, as depicted in Figure~\ref{fig:simplelikeslice} for the beam and noise of a 100 GHz detector. 
 We can quantify the bias in the likelihood by computing the distorted mean,
\begin{equation}
\bar n_s(\Delta) =   \int dn_s  \ n_s\  {\cal L} \left( {\cal D}_l(\Delta) | {\cal C}_l(n_s) \right),
\end{equation}
{plotting the distance from the true value as a function of $\Delta$}, normalized by the error in $n_s$ along that slice, given by 
\begin{equation}
  \sigma^2_{n_s} = \int dn_s \ (n_s - n_{s,{\rm true}})^2 \ {\cal L} \left( {\cal D}_l | {\cal C}_l(n_s) \right).
\end{equation}
Figure~\ref{fig:simplelikebias} summarizes the likelihood bias for a detector in each of the channels, when the fitted beam is too small.  The highest frequency channels (545 and 857 GHz) show no bias simply because the errors are so large, and are not displayed.  The pivot point for the family of spectra in our slice is $l \sim 570$,  so the channels (30 and 44 GHz) with larger beams, which weight lower $l$ more strongly, show a negative bias in $n_s$, while the others show a positive bias.  For the three channels with the best noise, the beam requirement is strictest, and the fidelity in the beam required is striking.  For example, at 100 GHz for a single detector, to limit the distortion in the likelihood slice to $0.1\sigma$, the FWHM must be known to 0.1\% for a symmetric Gaussian.  Note that our parametric fits (Table~\ref{tab:vanillaparameters}) are achieving this precision in all nine channels.  For a general beam, this means that the beam {window} function $b_l^2$ must be known to almost 0.04\% where it has fallen to 1 percent, and better at lower multipoles. 

\subsection{Realistic beams}

We can marginalize the likelihood over the errors in the {window} function,
\begin{equation}
\int d\{r_l\}\ {\cal L} \left( r_l {\cal D}_l | {\cal C}_l(n_s) \right) P(r_l | \mbox{ planet observation } ),
\end{equation}
using the {window} function ratio $r_l$ from Eq.~\ref{eqn:transrat}.  Assuming a flat prior on the {window} function, we may approximate the posterior probability distribution of {window} functions with our Monte Carlo ensemble of beam {window} functions.

We find the fidelity of the reconstruction for the parametric model is very impressive.  In the bands most important for the CMB, marginalizing over the ensemble increases the standard deviation of the likelihood slice for $n_s$ by 6\% at 100 GHz, 1\% at 143 GHz, and leaves the errors at 217 GHz essentially unchanged.  
We can also examine how the peak of the likelihood slice is shifted around for particular realizations in the beam-fitting ensemble.  For the parametric model, the rms peak shift is $0.6\sigma$ at 100 GHz, $0.3\sigma$ at 143 GHz, and $0.1\sigma$ at 217 GHz.

The impact on the errors is more substantial in the non-parametric beam decomposition.  The width of the likelihood slice for $n_s$ is increased by 11\% at 100 GHz; 9\% at 143 GHz; and 60\% at 217 GHz.  The ensemble rms bias in the peak of the likelihood is $1.2\sigma$ at 100 GHz, $1.1\sigma$ at 143 GHz, and $1.7\sigma$ at 217 GHz, indicating that the beam error is significant.

\section{Conclusions} \label{sec:conclusions}

We have examined the problem of fitting beams for \Planck\ to planet observations.  Using a simple, but rigid, parametric model, and a very flexible non-parametric model for the beam, we predict errors in the beam reconstruction from the focal-plane transit data with Monte Carlo simulations. As part of the development of the non-parametric beam decomposition, we showed how to evaluate the impact of a given scan strategy on the quality of beam reconstruction.  We note that elliptical Gaussian approximations fit to the simulated beams in real space produce substantial errors in the {window} functions, and should not be used for cosmological analysis.

The errors are much smaller in the parametric model, but it is a toy model holding the place for a detailed optical reconstruction of the telescope, which would probably require more parameters and provide less fidelity.  The non-parametric model depends only on the data from planet scans, and requires no modeling of the telescope optics.  Taking this as a pessimistic scenario for beam uncertainty, we project that a single transit of Jupiter should constrain the beam {window} function in the key 100-217 GHz CMB channels to 0.3\%.  This level of beam errors, however, will be a significant systematic for the measurement of the scalar spectral perturbation index $n_s$ as determined by a slice through the cosmological parameter likelihood.  Other sources of uncertainty in the planet measurement, such as a calibration error or low frequency structure in the detector response, will raise the uncertainty in the final cosmological measurement.

In this analysis, we have for simplicity excluded several effects which may prove important for the correct reconstruction of the beam.  These include uncertainties in the solution to the telescope's pointing, uncertainty in the planet's microwave frequency spectrum, relating to the differing sizes of the beams for planets and for the CMB, time variability in the planet signal (including the Galilean satellites of Jupiter, which by themselves will be high signal-to-noise signals for \Planck), and the finite size of the planet's disk.

\section*{Acknowledgments}
We thank Andrea Catalano, Fabio Noviello, Maura Sandri, and Vladimir Yurchenko for providing realistic models of the \Planck\ beams.  We also thank Ludovic Montier for providing HFI non-linear response curves.  Jean-Loup Puget, Jean-Michel Lamarre,  Francois Bouchet, and many members of the \Planck\ HFI and LFI Core teams gave helpful comments on this work.  Chris Hirata and Hans Kristian Eriksen provided useful discussions on the process of beam reconstruction.  
Jan Tauber, Marco Bersanelli, and Warren Holmes gave very useful comments on drafts of this paper.
Some results in this paper {made} use of the HEALPix software package \citep{2005ApJ...622..759G}.
This research used resources of the National Energy Research Scientific Computing Center, which is supported by the Office of Science of the U.S. Department of Energy under Contract No. DE-AC02-05CH11231. 
 This work was partially performed at the Jet Propulsion Laboratory,
 California Institute of Technology, under a contract with NASA.
KMH receives support from JPL subcontract 1363745.

\bibliography{planck_beam_fitting}
\bibliographystyle{aa}  

\end{document}